%% file: acl_latex.tex
\pdfoutput=1
\documentclass[11pt]{article}

\PassOptionsToPackage{hyphens}{url}\usepackage{hyperref}
\usepackage[final]{acl}

\usepackage{times}
\usepackage{latexsym}

\usepackage[T1]{fontenc}
\usepackage[utf8]{inputenc}

\usepackage{microtype}

\usepackage{inconsolata}

\input{config}

\title{Virtual Compiler Is All You Need For Assembly Code Search}

\author{
Zeyu Gao\textsuperscript{1},  % $\diamondsuit$}, 
Hao Wang\textsuperscript{1}, % $\diamondsuit$},
Yuanda Wang\textsuperscript{2}, and
Chao Zhang\textsuperscript{1}\thanks{corresponding author}\\
\textsuperscript{1}Tsinghua University \\ 
\textsuperscript{2}Beijing University of Posts and Telecommunications \\
{\tt \{gaozy22,hao-wang20\}@mails.tsinghua.edu.cn} \\
{\tt wangyuanda@bupt.edu.cn} \\
{\tt chaoz@tsinghua.edu.cn} \\ % 
}

\begin{document}
\maketitle

\input{body}

\clearpage
\newpage

\bibliography{Library,example_paper}

\appendix

\clearpage
\newpage

\input{appendix}

\end{document}

%% file: config.tex
\usepackage{multirow}
\usepackage{makecell}

\usepackage{inconsolata}
\usepackage{xspace}
\newcommand{\sysname}{{\tt ViC}\xspace}
\newcommand{\docstring}{{\tt UbuntuDocstring}\xspace}
\newcommand{\virtualdocstring}{{\tt VirtualDocstring}\xspace}
\newcommand{\vicds}{{\tt VirtualAssembly}\xspace}
\newcommand{\dsname}{{\tt HybridAssembly}\xspace}

\newcommand{\authnote}[2]{{\bf \textcolor{blue}{#1}: \em {\color{red}{#2}}}}
 
\newcommand{\wh}[1]{\authnote{wh}{#1}}

\usepackage{amsmath}
\usepackage{graphicx}

\usepackage{listings}

\usepackage{boxedminipage}
\usepackage{verbatim}

\usepackage{enumitem}

\usepackage[normalem]{ulem}
\useunder{\uline}{\ul}{}
\definecolor{shadecolor}{RGB}{246,246,246}

\definecolor{jade}{rgb}{0.0, 0.66, 0.42}
\definecolor{carolinablue}{rgb}{0.6, 0.73, 0.89}
\definecolor{dkgreen}{rgb}{0,0.6,0}
\definecolor{dkblue}{rgb}{0,0.4,0.5}
\definecolor{gray}{rgb}{0.5,0.5,0.5}
\definecolor{mauve}{rgb}{0.58,0,0.82}

\definecolor{codegreen}{rgb}{0,0.6,0}
\definecolor{codegray}{rgb}{0.5,0.5,0.5}
\definecolor{codepurple}{rgb}{0.58,0,0.82}
\definecolor{codeblue}{rgb}{0,0,205}
\definecolor{backcolour}{rgb}{245,245,245}

%% file: body.tex
\begin{abstract}
% Assembly code search is crucial in significantly reducing the workload of reverse engineers by enabling them to swiftly locate specific functionality from tens of thousands within binary files.
% Despite its significance, the task faces considerable challenges due to the difficulty in constructing high-quality datasets.
% This paper proposes a method of training a Large Language Model (LLM) to emulate a general compiler, resulting in \textbf{a virtual compiler capable of compiling multiple programming languages}. 
% This addresses the issue of compiling difficulties when constructing assembly code datasets.
Assembly code search is vital for reducing the burden on reverse engineers, allowing them to quickly identify specific functions using natural language within vast binary programs.
Despite its significance, this critical task is impeded by the complexities involved in building high-quality datasets.
\iffalse
This paper explores training a Large Language Model (LLM) to emulate a general compiler, leading to \textbf{a virtual compiler with the capability to compile multiple programming languages.}
\wh{multiple or any?} 
This reduces the labor-intensive compilation challenge, enabling transfer learning from high-level language code search to assembly code search.
We use the packages from Ubuntu to obtain a dataset with 20B tokens of source and assembly code, further continue pre-training CodeLlama into a \underline{\texttt{Vi}}rtual \underline{\texttt{C}}ompiler (\sysname).
It has the ability to compile across a wide range of languages without the real compiler,
facilitating a "virtual" compilation process retaining semantic equivalency, thereby broadening the scope for assembly code dataset construction.
\fi
This paper explores training a Large Language Model (LLM) to emulate a general compiler. By leveraging Ubuntu packages to compile a dataset of 20 billion tokens, we further continue pre-train CodeLlama as a \underline{\texttt{Vi}}rtual \underline{\texttt{C}}ompiler (\sysname), capable of \textbf{compiling any source code of any language to assembly code}. This approach allows for \textit{virtual compilation} across a wide range of programming languages without the need for a real compiler, preserving semantic equivalency and expanding the possibilities for assembly code dataset construction.
% Remarkably, this virtual compiler can compile not just C/C++ but also other high-level languages without native compilers, thereby broadening the scope for assembly code dataset construction.
% many of which lack real compilers, 
% To this end, 
Furthermore, we use \sysname to construct a sufficiently large dataset for assembly code search. 
Employing this extensive dataset, we achieve a substantial improvement in assembly code search performance, with our model surpassing the leading baseline by 26\%.
% The model trained on our constructed assembly dataset outperforms the best general embedding model by 26\% in assembly code search.
\end{abstract}

\section{Introduction}

% Reverse engineering processes are often encumbered with the formidable task of promptly pinpointing specific functions from tens of thousands within binary files, such as those associated with malicious behavior.
Reverse engineering frequently entails the challenging task of swiftly locating specific functions within extensive binary files, such as those associated with malicious behavior.
Traditionally, reverse engineers rely on the tedious approach of searching for unique strings~\cite{ida-strings-windows} or constants~\cite{findcrypt-yara} to locate these functions. 
% This practice, heavily reliant on experience and heuristics, frequently results in inefficient procedures where unidentifiable functions necessitate sequential or heuristic-based searching, consuming a substantial amount of time.
This practice often leads to inefficiency as it heavily relies on experience or heuristic algorithms, frequently consuming considerable time.

We observe that the specific search requirements articulated by reverse engineers could often be distilled into natural language descriptions. This realization sparked the idea of employing a code search~\cite{luCodeXGLUEMachineLearning2021,fengCodeBERTPreTrainedModel2020} perspective to refine the extant methodologies. Code search has achieved substantial success in high-level programming languages, garnering wide commercial application~\cite{github-towards-natural-language-semantic-code-search}.
Assembly code search enables users to search the assembly (i.e. assembly code) function with natural language, enhancing user interaction with binary executable files through natural language.

The construction of assembly code search datasets presents considerable difficulty. Unlike high-level programming languages where datasets can be readily created by parsing the docstring of function, constructing an assembly code search dataset entails an obligatory compilation step from source code to assembly code. It is exemplified by the work conducted in Nova$^+$~\cite{jiangNovaGenerativeLanguage2023}. Despite attempts to compile over 4 million C programs and 1 million C++ programs, only 32,774 C programs and 40,087 C++ programs were successfully compiled. 
Numerous compilation failures stem from the complex compilation environment, including diverse dependencies and variations in optimization levels and compilers, greatly complicating dataset construction.
% Previously some works aimed at addressing compilation issues of single-source C (Without C++ support) functions using type inference, however, due to the scarcity of code search datasets in C source, the assembly code search dataset cannot be augmented in this way easily.
Previous efforts target the compilation challenges of single-source C functions through type inference or neural compilation~\cite{dasilvaAnghaBenchSuiteOne2021,guoEnablingTransformersUnderstand2022,armengol-estapeLearningX86Translation2021}. Yet, the practical utility of these advancements is constrained by the scarce availability of code search datasets for C language.

Inspired by Meta's use of Large Language Models (LLMs) for compiler optimization~\cite{cumminsLargeLanguageModels2023},
% we postulate a novel modeling paradigm—compiling directly from source code to corresponding assembly instructions, allowing LLMs to emulate the compiler and assimilate the comprehensive compilation process, including optimization options and assembly code generation.
we postulate a novel modeling paradigm—let LLMs emulate a general compiler and understand the comprehensive compilation process, including optimization options and assembly code generation.
This method not only enables us to utilize the previous code search dataset in C/C++ language like CCSD~\cite{liuRetrievalAugmentedGenerationCode2021} but also demonstrates the capability to generalize across languages such as Python and Golang, despite only being trained on C/C++ source and assembly code pairs.
This flexibility suggests the potential of applying extensive datasets from prior works, like CodeSearchNet~\cite{husainCodeSearchNetChallengeEvaluating2020}, to the realm of assembly code search, addressing the historical constraints of dataset scarcity in this field.

% This virtual compiler can utilize the cleaned datasets from previous code search studies like CCSD~\cite{liuRetrievalAugmentedGenerationCode2021}, yielding an unbiased compilation of individual functions with remarkably promising results.

% In summary, our data-centric approach led to the formulation of an efficient schema tailored for binary code search datasets, laying the groundwork for binary code search advancements. Coupled with a robust pre-trained model, we embarked on training specifically tailored for binary code search, steering the understanding of binary programs in new directions beyond traditional binary code similarity detection (BCSD).

% In light of the challenges present in assembly code search and the limitations of existing methods, our work culminates in the following contributions:
Addressing the challenges in assembly code search and the shortcomings of current methodologies, our work makes the following contributions:

\begin{itemize}
    \item \textbf{Introduction of the virtual compiler (\sysname)}.
    We introduce a novel approach, \sysname for creating an assembly code search dataset. By employing the virtual compiler, we generate a diverse and robust dataset that circumvents the traditional barriers related to compilation challenges, vastly enriching the resources available for assembly code search tasks. % To our knowledge, we are the first to propose the concept of a virtual compiler.
    \item \textbf{Enhanced assembly code search performance}.
    We constructed a high-quality assembly code dataset using a virtual compiler, and the model trained on this dataset achieved a 26\% performance improvement in assembly code search tasks over existing state-of-the-art (SOTA) solutions. 
    \item We release our models and datasets to facilitate future research\footnote{\href{https://github.com/zeyugao/VirtualCompiler}{https://github.com/zeyugao/VirtualCompiler}}.
\end{itemize}

\section{Background and Related Works}

\subsection{Assembly Code Analysis}

\begin{figure}[t]
    \centering
    \includegraphics[width=1\linewidth]{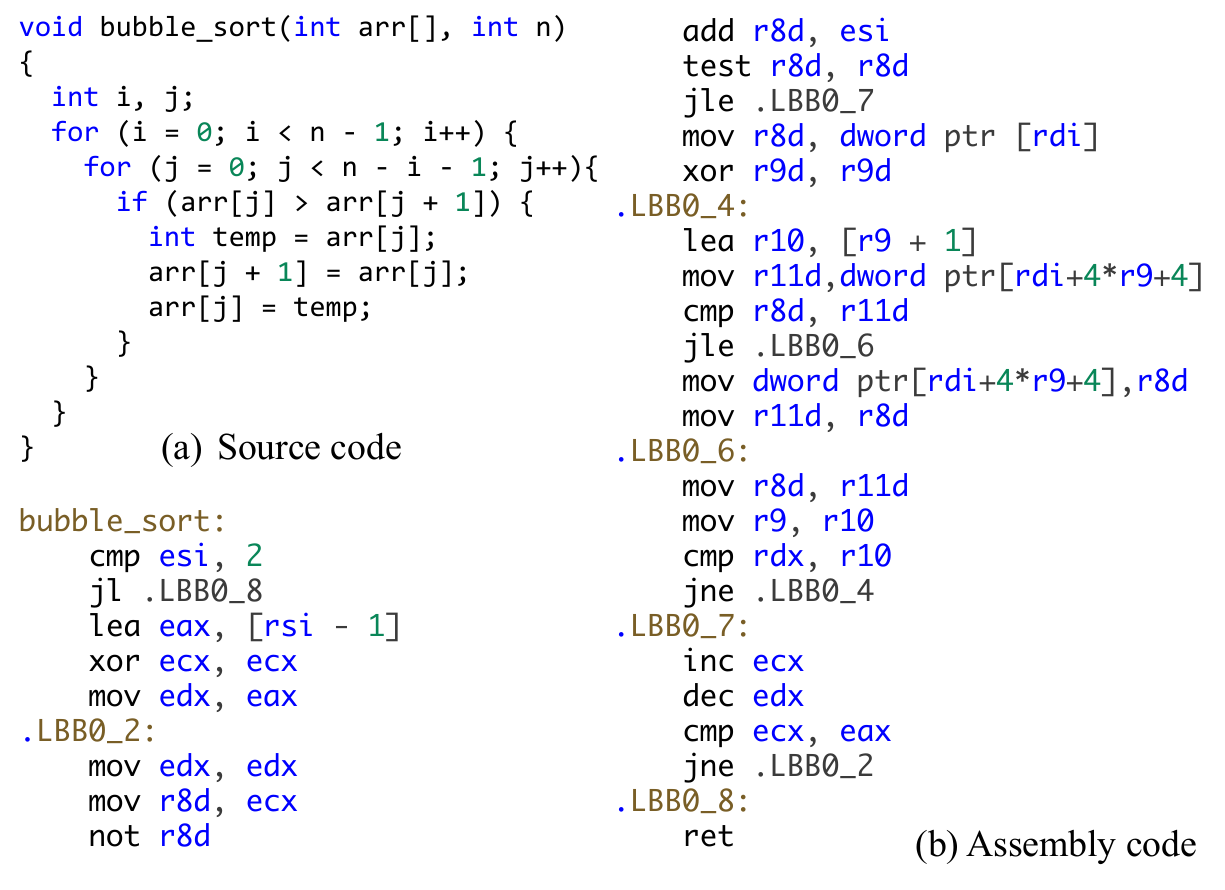}
    \caption{C source code and the compiled assembly code for a bubble sort algorithm.}
    \label{fig:bubble-sort}
    % \vspace{-2mm}
\end{figure}

The process of transforming high-level, human-readable source code (such as C or C++) into assembly code (also known as binary code) that a CPU can execute directly is known as compilation. Figure~\ref{fig:bubble-sort} shows the source code and the corresponding assembly code for the bubble sort algorithm.
It clearly illustrates that throughout compilation, much of the original context and structure, including variable names and high-level logical structures like loops and conditional statements, are lost.
This loss makes understanding compiled binary challenging, which is crucial for identifying security vulnerabilities and patching bugs, necessitating binary code analysis. 

Binary reverse engineering is a commonly used methodology in binary code analysis. It entails analyzing binary files without the source code to reconstruct the function functionality and partial program logic. In this complex process, one of the challenges is locating the specific function with certain functionality, such as encryption and authentication. The common methods include searching for representative strings~\cite{ida-strings-windows}, analyzing function import and export, looking for pattern matching~\cite{findcrypt-yara}, static analysis and dynamic debugging. These methods, while useful, can be time-consuming and imprecise. To overcome these limitations, the development of assembly code search models marks a significant advancement, offering a more efficient and accurate method of navigating binary files.

\subsection{Assembly Code Modeling}
With high-level logical structure replaced by the bare jump instructions, understanding and modeling the assembly code has quite different challenges compared to the high-level language.
% , such as C++ and Rust.
To effectively address these challenges, researchers have developed various techniques to model the assembly code for the downstream tasks such as binary code similarity detection (BCSD). This technique is pivotal for assessing the similarity between different assembly codes which can be affected by varying aspects like compiler versions and optimization levels and identifying the paired assembly code compiled from the same source code from an amount of candidates, useful for clone detection and supply chain analysis.
%, representing binary code as continuous vectors in a vector space, suitable for tasks such as binary code similarity detection (BCSD) and reverse engineering. 
% MalConv~\cite{raffMalwareDetectionEating2017} and $\alpha$-Diff~\cite{liuDiffCrossVersionBinary2018}, which employ CNNs and LSTMs to analyze binary code from raw bytes.
% DeepDi~\cite{DeepDiLearningRelational2022} constructs an instruction flow graph and utilizes a relational graph convolutional network for precise classification of instruction vectors. 
CodeCMR~\cite{yuCodeCMRCrossModalRetrieval2020} integrates GNN, DPCNN, and LSTM to extract the semantic features of source code and binary code.
PalmTree~\cite{liPalmTreeLearningAssembly2021} pre-trains on unlabeled binary corpus through self-supervised training to capture various characteristics of assembly languages.
Trex~\cite{peiTrexLearningExecution2021} models micro traces with Transformers, while jTrans~\cite{wangJTransJumpawareTransformer2022} uses unique jump-aware representation method preserves control flow information.
These models, especially the Transformer-based models, have a partial ability to capture the semantics in the function to support the code search task.

% \subsection{Source Code Compilation}

\subsection{Code Search}

\begin{figure*}[t]
    \centering
    \setlength{\abovecaptionskip}{2mm}
    \includegraphics[width=\linewidth]{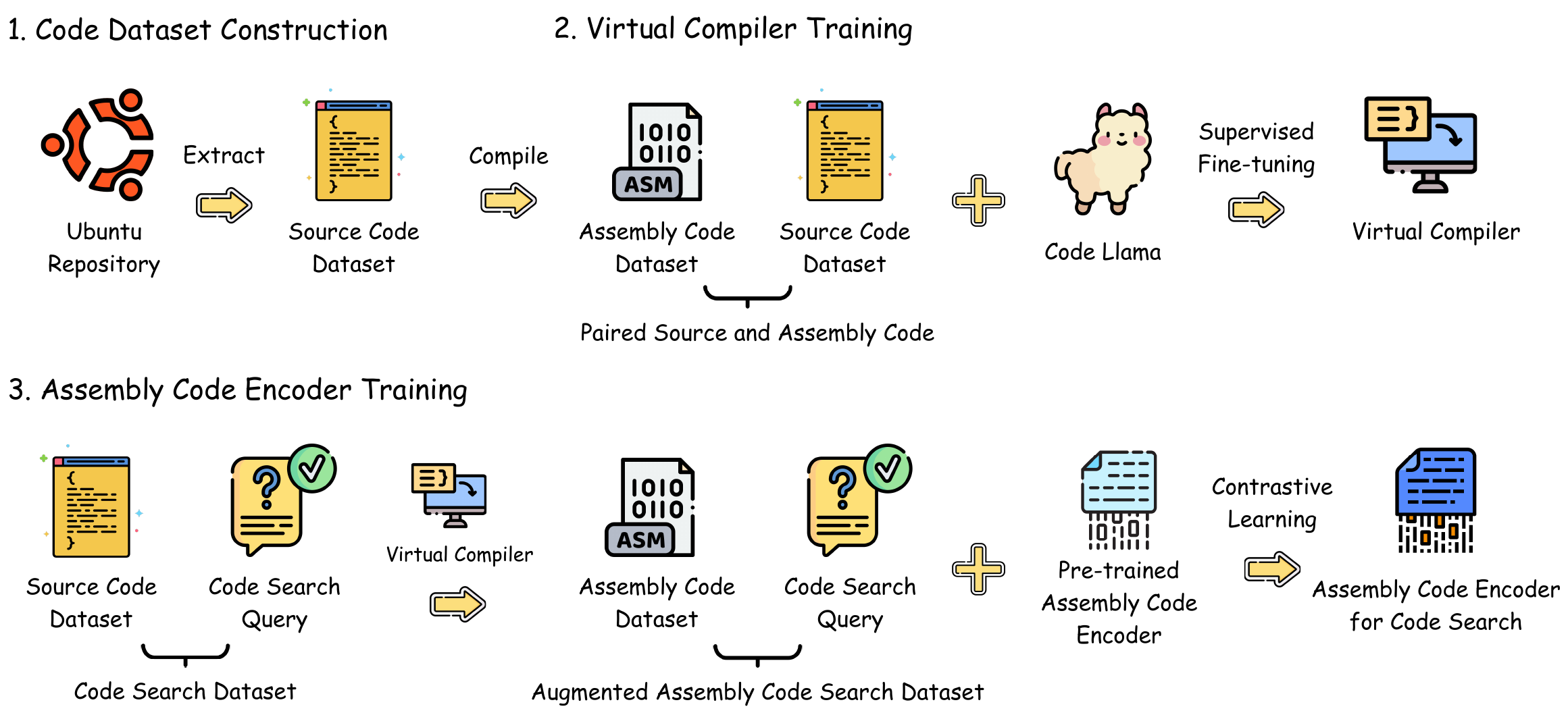}
    \caption{The workflow overview of using \sysname for assembly code search.}
    \label{fig:vic-overview}
    % \vspace{-2mm}
\end{figure*}

% Semantic code search refers to the process of querying and retrieving code snippets or software artifacts based on their semantic content rather than solely on syntactic matches. 
% This advanced form of code search has significant implications in software engineering and reverse engineering, where the intent behind a query often involves complex functionalities or concepts that cannot be adequately captured by keywords alone. Semantic code search leverages the understanding of the code's underlying purpose, enabling more accurate and relevant search results.

Code search plays a pivotal role in software development, enabling developers to query and retrieve valuable code snippets. This process leverages either pattern matching or, more recently, natural language queries, thanks to advancements in semantic code search. The most code search works hinges on a pipeline that begins with pre-training models on extensive corpora consisting of unpaired code and natural language and accompanies by fine-tuning the model on specialized code search datasets, which sharpens their ability to understand and execute code search queries accurately~\cite{wangCodeT5OpenCode2023,wangCodeT5IdentifierawareUnified2021,luCodeXGLUEMachineLearning2021,fengCodeBERTPreTrainedModel2020,guoGraphCodeBERTPretrainingCode2021}.

% With the emergence of Large Language Models, code search has gained prominence as a crucial component within Retrieval-Augmented Generation (RAG)~\cite{langchain,gaoRetrievalAugmentedGenerationLarge2023}. The RAG module leverages the code search module to fetch relevant code snippets based on the context for more accurate answer generation and assembly code search is crucial for reverse engineering.

% \subsection{Large Language Model}

\section{Overview}

% We present our workflow in Figure~\ref{fig:vic-overview}. In Section~\ref{sec:virtual-compiler}, we introduce the first and second steps shown in the overview, giving the details about how we train the Virtual Compiler. In Section~\ref{sec:code-search-contrastive-learning}, we show how we use Virtual Compiler to bootstrap the binary code search and the details in contrastive learning we use.

We present our workflow in Figure~\ref{fig:vic-overview}, with three phases to advance assembly code search capabilities. Initially, we compile a vast collection of packages from Ubuntu to obtain a mapping between source and assembly code. Following this, we train the CodeLlama model on this massive dataset to emulate compiler behavior (Section~\ref{sec:virtual-compiler}). Lastly, leveraging this model, we perform virtual compilations on existing code search datasets to produce an augmented assembly code search dataset. 
This enriched dataset then informs the training of an assembly code encoder, yielding a refined model for effective assembly code search (Section~\ref{sec:code-search-contrastive-learning}).

\section{Virtual Compiler}
\label{sec:virtual-compiler}

\subsection{Code Dataset Construction}
\label{subsec:data-processing}

In our pursuit to train a model to compile the source code of the individual function to its corresponding assembly code, we compile over 6,000 C/C++ packages from Ubuntu using two types of x86-64 compilers, GCC and Clang, each in three different versions (\texttt{GCC-\{7,9,11\}} and \texttt{Clang-\{9,11,12\}}) and five optimization levels (\texttt{O\{0-3\}} and \texttt{Os}). 
% It is worth noting that changing the compiler version and optimization level will likely make compiling fail. 
Then, we extract the source and assembly code for the model training. The model exclusively provides the source code of the function body and is required to predict the corresponding assembly code.
To reduce the hallucinations of the model, we exclude assembly code containing inline functions due to compiling optimization to ensure a one-to-one correspondence between the source code and assembly.
Additionally, we omit functions shorter than 5 lines. These often represent class constructors, which will generate disproportionately large volumes of assembly code.
% Also, we exclude functions having fewer than 5 lines in the source code as there is a large possibility that the source code is a constructor of a class, resulting much much more assembly code, leading to huge hallucinations of the model. 
However, for simplicity in source code processing, global variables, and macro definitions are not provided to the model.

Ultimately, we amass a dataset consisting of 15 million source-to-assembly pairs, exceeding 20 billion tokens. We filter out source functions with fewer than four lines in the function body, remove function comments, and prefer unmangled function names in the assembly code.
%, noting that the model struggles with accurately predicting mangled names.
A validation set of 1,000 unseen function source codes paired with corresponding assembly code is segregated to assess the output quality.

% \subsection{Prompt}

% The model is trained following the Supervised Fine-Tuning (SFT) method, where the System Message and User Input include instructions for the model and the function source code, and the Assistant Output consists of the resultant assembly code. During the loss calculation, the loss is computed solely based on the Assistant Output, excluding the instructions and function source code. Instructions include the compiler type, optimization level, and whether the assembly code should be stripped, denoting the removal of symbol information, such as function name and local variable name.

\subsection{Model Training}

Contrasting with Meta's work~\cite{cumminsLargeLanguageModels2023} that begins with random initialized weight, we employ Codellama 34B~\cite{roziereCodeLlamaOpen2023} as our foundational model, anticipating a more rapid convergence~\cite{chenEvaluatingLargeLanguage2021}. Subsequent experiment shows that leveraging Codellama also provides enhanced comprehension of various programming languages. 
The model is trained using Supervised Fine-Tuning (SFT). Besides the function source code, the prompt includes the compiler, optimization level, and whether the assembly code should be stripped, denoting the removal of symbol information, such as local variable names.

% Considering the input length in the CodeSearchNet dataset, despite Codellama supporting up to 16k tokens, we only use 4k tokens as context length for faster training.
% We confine our training to source-to-assembly code pairs under 4k tokens for faster training, accommodating a substantial portion of longer training instances from the CodeSearchNet dataset.
Considering the length of the vast majority of inputs in the CodeSearchNet, we opt for a 4,096 context length during training instead of the 16,384 natively supported by CodeLlama, aiming for a faster training process.
We 
% embrace the Curriculum Learning strategy~\cite{curriculum-learning}, 
initially focus on training sets with less than 2,048 tokens, followed by a training mix that includes a 3:1 ratio of 0-2,048 and 2,048-4,096 token-length data for better training efficiency. The model is trained on 64 NVIDIA A100 GPUs, exceeding a total of 1,000 GPU days, employing a cosine learning rate schedule with 1\% of warm-up steps. The initial phase operates at a peak learning rate of 3e-5 with a batch size of 3,072, while the subsequent phase uses a peak learning rate of 1.5e-5 and a batch size of 2,048. Each dataset underwent a single epoch.

\section{Code Search Contrastive Learning}
\label{sec:code-search-contrastive-learning}

% In this section, we delineate the training methodology adopted for aligning the binary code model with its code search textual representations.

\subsection{Dataset}
\label{subsec:constrastive-learning-dataset}

% e discuss our innovative approach to utilizing datasets from diverse languages or projects that are incompilable in their original form. Through the process of virtual compilation, we synthesized a training dataset conducive to code search contrastive learning.

We use two datasets to conduct the code search contrastive learning.
The first dataset we utilize is CCSD~\cite{liuRetrievalAugmentedGenerationCode2021}, which comprises function summarization from over 95,000 functions from 300 different projects with an extensive deduplication and cleaning process.
% These functions, which lack associated project metadata such as version information, cannot be traced back to their source for traditional compilation.
Furthermore, we apply virtual compilation to CodeSearchNet~\cite{husainCodeSearchNetChallengeEvaluating2020}, a challenge introduced by GitHub featuring a dataset and a corresponding benchmark. This dataset is the bedrock of contemporary code search research nowadays.
% , directly utilized by projects such as GTE~\cite{liGeneralTextEmbeddings2023}  and sentence-transformers~\cite{reimersSentenceBERTSentenceEmbeddings2019} for code search training. 
CodeSearchNet spans multiple programming languages — Python, JavaScript, Ruby, Go, Java, and PHP — none of which are directly applicable to assembly code search due to their higher-level nature.

In constructing our virtual assembly dataset for model training, each function from CCSD and CodeSearchNet is subjected to a randomly chosen combination of compiler and optimization levels, thus generating their virtual assembly codes. This approach allows us to expand the applications of these comprehensive, multi-language datasets into the realm of assembly code search.

\subsection{Model Architecture}

Building on the revelations from jTrans~\cite{wangJTransJumpawareTransformer2022}, we acknowledge the need for specialized treatment of textual and assembly code representations. To this end, we decouple the text encoder and assembly code encoder, adopting a bespoke architecture for the assembly code encoder, allowing each encoder to become adept at handling the intricacies of its respective data modality. 

For the assembly code encoder, we use a roformer-base~\cite{suRoFormerEnhancedTransformer2022} model that incorporates shared parameters~\cite{wangJTransJumpawareTransformer2022} with 110M parameters to better integrate the inductive bias of control flow in assembly code. And we keep string literals in assembly code to better preserve semantic information through Byte-Pair Encoding instead of normalizing them, which is a common simplification in previous assembly modeling work~\cite{liPalmTreeLearningAssembly2021,peiTrexLearningExecution2021,wangJTransJumpawareTransformer2022}. On the other side, the text encoder is initialized by the well-established sentence-transformers~\cite{reimersSentenceBERTSentenceEmbeddings2019}.
% Given the sentence-transformers's proficiency in capturing the textual nuances in code search scenarios, we freeze the text encoder when training. Instead, we opt to align the binary code encoder with the vector space of the pre-existing, well-tuned text encoder, tapping into the latent semantic understanding already etched into the GTE model.

\subsection{Assembly Code Encoder Training}
\label{subsec:assembly-code-encoder-traininig}

During the training of the assembly code encoder, we employ the commonly used in-batch negative sampling strategy and use InfoNCE loss as the training loss, shown as Equation~\ref{eq:in-batch-negative-1} and~\ref{eq:in-batch-negative-2}. This method aims to increase the similarity between positive pairs within a batch while reducing the similarity across negative pairs. The text encoder is initialized from sentence-transformers, and the assembly encoder is pre-trained with MLM (Masked Language Model) and JTP (Jump Target Prediction) tasks. With the text encoder well-established, we set the learning rate to 1e-6 to reduce the disturbance, and the learning rate for assembly code is set to 2e-5 empirically~\cite{wangJTransJumpawareTransformer2022}.
We take the model with the best in-dataset evaluation result for further evaluation.

% Throughout the alignment procedure, we considered two distinct loss functions for training. The first loss comes from CLAP, which addresses the challenge of a binary code differentiating the matching text from a set of $N$ possible texts. Due to the text encoder being frozen, we can afford to push the boundaries of $N$, hence enlisting an extensive text for the binary code to compare with. 
% The second loss function embraces the concept of in-batch negative sample loss. Within the confines of a single batch, each binary code is matched with text in a mutual search process. We optimize the InfoNCE loss of these two targets.

Give a batch of positive assembly code and text pairs $B = \{(a_1, t_1,), (a_2, t_2), \cdots, (a_n, t_n)\}$. We treat each pair $(a_i, t_i)$ as positive pair, and $(a_i, i_j)_{i \neq j}$ as negative pairs. The text-to-assembly contrast loss is defined as:

\begin{equation}
    % L_1 = \frac{1}{n} \sum_i^n L_E(a_i, t_i; q_1, q_2, \cdots, q_N)
    % \label{eq:large-text-neg-loss}
    L_1 = - \frac{1}{n} \sum_{i=0}^{n} \log \frac{\exp(t_i \cdot a_i / T)}{   \sum_{j=1}^{N} \exp(t_i \cdot a_j / T)   }
    \label{eq:in-batch-negative-1}
\end{equation}

The auxiliary loss, assembly to text contrast loss, is inversely defined as: 

\begin{equation}
    % L_1 = \frac{1}{n} \sum_i^n L_E(a_i, t_i; q_1, q_2, \cdots, q_N)
    % \label{eq:large-text-neg-loss}
    L_2 = - \frac{1}{n} \sum_{i=0}^{n} \log \frac{\exp(a_i \cdot t_i / T)}{   \sum_{j=1}^{N} \exp(a_i \cdot t_j / T)   }
    \label{eq:in-batch-negative-2}
\end{equation}

where $T$ is the temperature, which we set to 0.07 empirically~\cite{radfordLearningTransferableVisual2021,heMomentumContrastUnsupervised2020}.

Then the train loss is defined as $L = L_1 + L_2$.

% Denote the InfoNCE loss function in Eq.~\ref{eq:infonceloss}, with $p, q$ as the positive pair, $\{k_1, k_2, \cdots, k_N\}$ as negative samples.

% \begin{align}
% \begin{split}
%     L_E(p, q; k_1, k_2, &\cdots, k_n) = \\
%     - \log &\frac{\exp(p \cdot q)}{\exp(p \cdot q)   + \sum_{j=1}^{N} \exp(p \cdot k_i)   }
%     \label{eq:infonceloss}
% \end{split}
% \end{align}

% The first extensive negative text loss is defined by Eq.~\ref{eq:large-text-neg-loss}

% \begin{equation}
%     L_1 = \frac{1}{n} \sum_i^n L_E(a_i, t_i; q_1, q_2, \cdots, q_N)
%     \label{eq:large-text-neg-loss}
% \end{equation}

% The second in-batch mutual search loss is defined by Eq.~\ref{eq:in-batch-mutual-search}

% \begin{align}
% \begin{split}
%     L_2 &= \frac{1}{n} \sum_i^n L_E(a_i, t_i; t_1, \cdots, t_{j, j\neq i}, \cdots, t_n) \\
%     &+ \frac{1}{n} \sum_i^n L_E(t_i, a_i; a_1, \cdots, a_{j, j\neq i}, \cdots, a_n)
%     \label{eq:in-batch-mutual-search}
% \end{split}
% \end{align}

\section{Evaluation}

\subsection{Evaluation Setup}

We implement \sysname using Pytorch 2.1~\cite{paszkePyTorchImperativeStyle2019}. We use IDA Pro 8.3~\cite{idapro} to disassemble and extract the functions from the binary executable file in all of the experiments. Our training and experiments are conducted on several servers to
accelerate training. The CPU setup is 128 cores with 2TB RAM for each server. The total GPU setup is 32 NVIDIA Tesla A100.

\subsection{Evaluation of Virtual Compiler}

We endeavor to appraise the quality of assembly code yielded by the virtual compiler, with an emphasis on the model's capability to generate assembly code for augmenting the code search dataset.

We employ two distinct approaches for evaluating the model's capability. For code with ground truth, namely, C/C++ source code with corresponding assembly code, we assess quality using various similarity metrics in Section~\ref{subsubsec:sequence-similarity}-\ref{subsubsec:semantic-similarity}. For high-level languages without direct assembly code ground truth, such as PHP and Golang from the CodeSearchNet dataset, we use the downstream task in Section~\ref{subsec:evaluation-assembly-code-search} and a case study in Section~\ref{sec:illustration-virtual-compiling-golang} to demonstrate the model's capabilities.

During the dataset construction phase, we preserve 1,000 C/C++ functions with their source code and assembly code that the model has not previously trained in the training process for evaluation.

\subsubsection{Sequence Similarity}
\label{subsubsec:sequence-similarity}

\begin{figure}[t]
    \centering
    \setlength{\abovecaptionskip}{2mm}
    \includegraphics[width=1\linewidth]{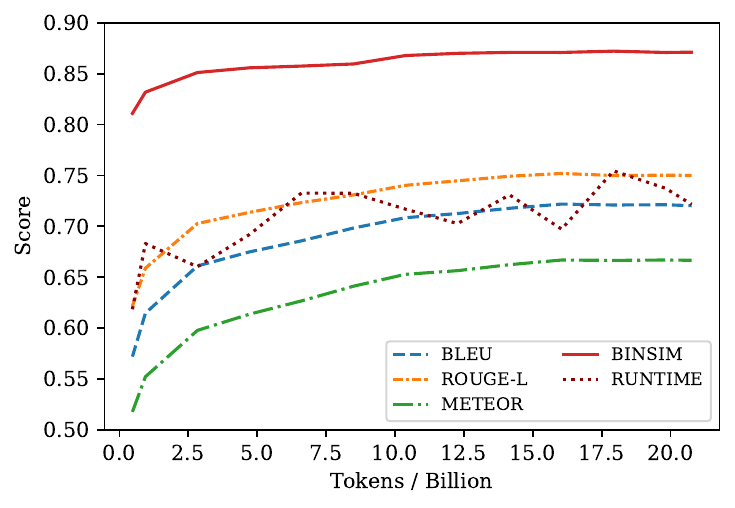}
    \caption{Correlation between the number of tokens used in model training and the quality of generated assembly code, as evaluated by various metrics}
    \label{fig:bleu-binsim}
    % \vspace{-2mm}
\end{figure}

A concise evaluation of the generated assembly code is carried out using the BLEU~\cite{papineniBleuMethodAutomatic2002}, ROUGE-L~\cite{linROUGEPackageAutomatic2004}, and METEOR score~\cite{banerjeeMETEORAutomaticMetric2005}, which are employed to show the similarity between generated assembly code and the reference ground truth.
The scores are shown in Figure~\ref{fig:bleu-binsim}, revealing a clear trend in the data that as the number of tokens processed during model training increased, there is a commensurate rise in the scores for each metric. 
% The final BLEU score surpassed the 0.7.

\subsubsection{Runtime Similarity}
\label{subsubsec:runtime-similarity}

We further study the runtime characteristics of the assembly code generated by the virtual compiler. We implement an enforced execution of assembly code instructions using Keystone~\cite{keystone} and Unicorn~\cite{unicorn}, executing any assembly code by randomly initializing the CPU's registers and memory state. 
% We execute assembly code for 1,000 functions generated by a compiler and assembly code generated by the virtual compiler, 
We execute assembly code generated by the real and virtual compilers, respectively, and record their memory reads and writes. To avoid infinite loops, we set a maximum executed instruction count of 2,000.

We compare the function return value, the local stack of the function, and the sequence of memory accesses as indicators for studying the runtime characteristics of functions. Specifically, we use a very strict evaluation scheme to compare whether the \texttt{rax} register, which holds the return value, is equal at the end of execution, whether the stack registers \texttt{rsp} and \texttt{rbp} are equal, and whether the sequence of memory reads and writes are equal. We denote the average of these three indicator as \texttt{RUNTIME} in Figure~\ref{fig:bleu-binsim}.
As the number of training tokens increases, the runtime characteristics of the virtual assembly become increasingly similar to those of the ground truth.

\subsubsection{Semantic Similarity}
\label{subsubsec:semantic-similarity}

Complementing the BLEU score, we adopt a more contextualized assessment method, Binary Code Similarity Detection (BCSD), aligned with our code search task.
% When the BCSD model reported high similarity scores, it implies that utilizing such assembly code interchangeably for a downstream task would likely produce similar outcomes, suggesting high substitutability. 
% This ability to exchange one assembly code version for another in downstream tasks without loss of performance is exactly what our task demands, highlighting binsim not only as a suitable metric but also as an integral part of our overall system's evaluative suite. 
In this task, we use CLAP~\cite{wang2024clap} for evaluation.
% \elsa{Try to fix the citation or declare the originate of the model}
As shown in Figure~\ref{fig:bleu-binsim}, after training on only 0.5 billion tokens, the virtual compiler is already capable of producing assembly code achieving a \texttt{BINSIM} score above 0.8, and it nears convergence after processing 5 billion tokens. In the context of BCSD tasks, a \texttt{BINSIM} score exceeding 0.8 typically indicates a strong positive pair of functions, with their variations likely attributable to different compilers or optimization options. This signifies that the assembly code generated by our model is perceived as being remarkably close to the ground truth.

% and implies that utilizing such assembly code interchangeably for a downstream task would likely produce similar outcomes, suggesting high substitutability. This ability to exchange one assembly code version for another in downstream tasks without loss of performance is exactly what our task demands, highlighting binsim not only as a suitable metric but also as an integral part of our overall system's evaluative suite.

\subsubsection{Case Study}

Given the complexity inherent in the virtual compilation process, perfect replication of compiler-generated results is a considerable challenge. This complexity is primarily due to the extensive presence of addresses, global variables, and the use of structure offsets in the code, which add intricate layers to the compilation task. While we utilize quantitative measures like BLEU to gauge the similarity between the model-generated and compiler-generated assembly code, we recognize that such metrics might not capture the full spectrum.

\begin{figure}[t]
    \centering
    \setlength{\abovecaptionskip}{2mm}
    \includegraphics[width=1\linewidth]{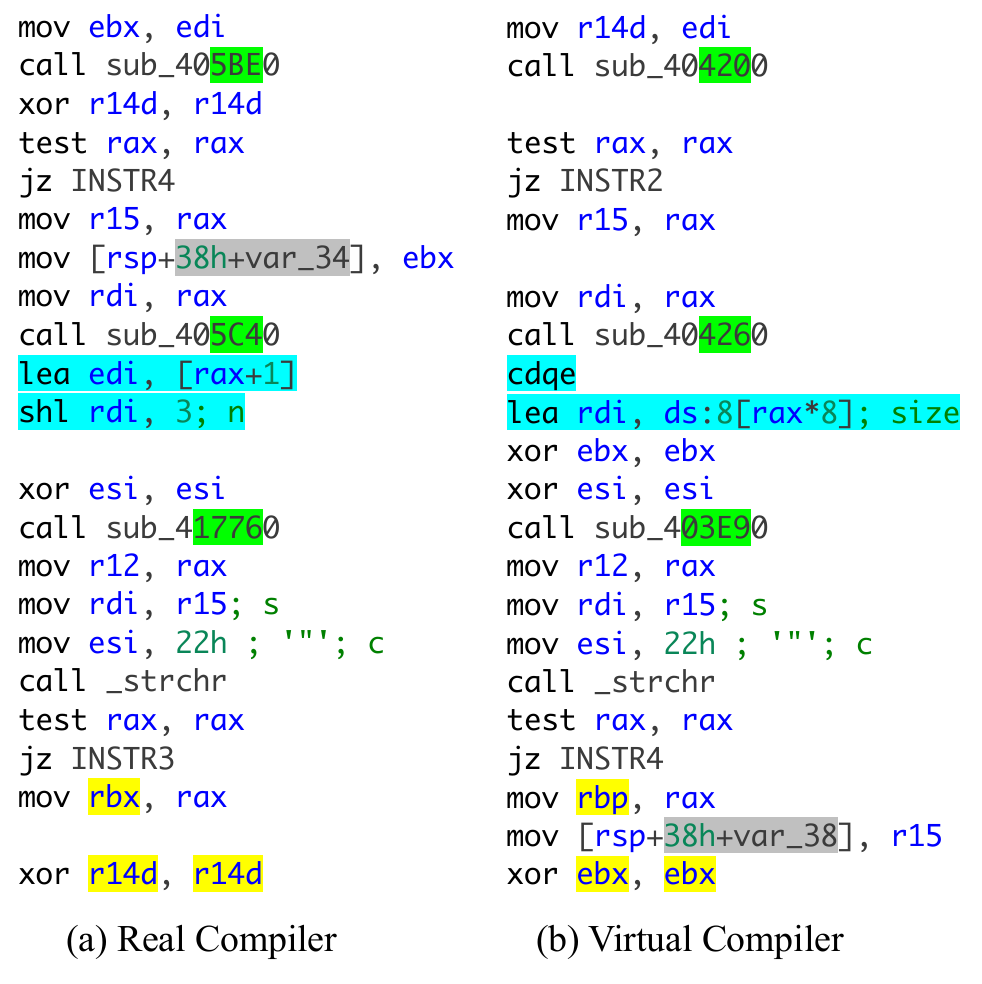}
    \caption{Comparison of assembly code from the (a) real compiler and (b) virtual compiler. Mismatches are highlighted in green (address differences), cyan (alternate operation expressions), yellow (register allocation variances), and grey (stack allocation discrepancies).}
    \label{fig:vic-compare-1}
    % \vspace{-2mm}
\end{figure}

In Figure~\ref{fig:vic-compare-1}, we present a segment of a large function to showcase the comparison between the model-generated assembly code and the real compiler-generated counterpart. A cursory glance reveals a multitude of differences. However, a thorough manual inspection reveals several distinct categories of mismatches, which we have highlighted in the figure using different colors for clarity.

The mismatches highlighted in green signify differences in the addresses. These discrepancies arise from the model's inability to obtain addresses during the virtual compilation.
% , leading it to make random guesses. 
Cyan highlights indicate alternative expressions of the same operation. For example, the ground truth assembly executes the operation \texttt{rdi = 8 * (rax+1)} using \texttt{lea edi, [rax+1]; shl rdi, 3}, whereas the virtual compiler opts for \texttt{lea rdi, ds:8[rax*8]}. Although semantically equivalent, such variations are not captured by string-matching metrics like BLEU.

The third category of mismatches, marked in yellow, corresponds to differences in register allocation during code generation. The actual compiler may choose a different set of registers compared to the virtual compiler. Grey highlights denote inaccuracies in stack allocation and usage due to the model's unawareness of stack variables' sizes.

Finally, an unresolved category of divergence, not explicitly marked in the figure, reveals additional instructions provided by one compiler that are absent in the other's output, and vice versa. This discrepancy underscores the inherent unpredictability and complexity when comparing outputs from two different compilers.

% \subsubsection{Cross-Language Assembly Compilation}

\subsection{Evaluation of Assembly Code Search}
\label{subsec:evaluation-assembly-code-search}

In this section, we validate the enhancement of the model's code search capabilities brought by the virtual compiler. Moreover, we aim to indirectly demonstrate the quality of the model-generated virtual assembly code and the model's capability to handle high-level languages.

% \subsection{Evaluation Setup}

\subsubsection{Training Dataset}

Similar to other code search datasets, we use the docstrings collected from the source code for comparison and apply cleaning steps in Section~\ref{sec:docstring-clean}.
% These docstrings are extracted along with the source code during the training of the \sysname, considering consecutive comment lines preceding a function's source code as the function's docstring. 
Out of 6,000 packages and 5 million source code functions, we extracted 400,000 docstrings, indicating that only a small fraction of functions are accompanied by docstrings.
We also use the virtual assembly code in Section~\ref{subsec:constrastive-learning-dataset}, including the code search dataset from  CodeSearchNet and CCSD. We show their statistics for each dataset in Table~\ref{tab:dataset-stat}. We denote the assembly code dataset generated by \sysname as \vicds, denote the extracted docstring dataset as \docstring, and denote the mixture of these two datasets as \dsname.
% For the dataset generated by the Virtual Compiler, we follow the methodology outlined in Section~\ref{subsec:constrastive-learning-dataset} to perform virtual compilation on two datasets, CodeSearchNet and CCSD, thereby obtaining their corresponding virtual assembly codes.

{\renewcommand{\arraystretch}{1.2}

\begin{table}[]
\small
\centering
\begin{tabular}{lccc}

\Xhline{1pt}
% Dataset & \begin{tabular}[c]{@{}c@{}}Pair Count\\ (K)\end{tabular} & \begin{tabular}[c]{@{}c@{}}Assembly Code\\ Tokens (M)\end{tabular} \\ \hline
Dataset       & Pair Count (K) & Asm Tokens (M) \\ \hline
CodeSearchNet & 1822.3 & 534.5          \\
\quad Go            & 280.5  & 66.8           \\
\quad Java          & 446.1  & 123.7          \\
\quad JavaScript    & 123.2  & 36.0           \\
\quad PHP           & 519.7  & 155.7          \\
\quad Python        & 404.2  & 137.8          \\
\quad Ruby          & 48.6   & 14.5           \\ \hline
CCSD          & 88.1   & 11.4           \\ \hline
Docstring     & 408.2  & 159.7 (5562.8) \\
\Xhline{1pt}

\end{tabular}
\caption{
\label{tab:dataset-stat}
The statistics for the dataset used for training.
Pair Count represents the count of query-assembly code pairs.
Asm Tokens represent the assembly code tokens. 
% In docstring dataset, 
Due to different compilers and different optimization levels used, we show the total augmented assembly code tokens in brackets for `Docstring'.
}
\end{table}

}

\subsubsection{Evaluation Dataset}

To construct the evaluation dataset of assembly code search that can represent the real-world scenario, we collect multiple popular binaries from the real world that have no overlap with our training set, from three different platforms: Mac, Windows, and Linux. Notably, although the model is trained using assembly code exclusively from Linux, the x86 instruction set assembly code has the same syntax across different operating systems, thereby enabling the model's cross-OS transferability.
Then we perform reverse engineering on these software applications. During this process, by leveraging the expertise of reverse engineering professionals, we write code search queries for functions that exhibit relatively independent functionalities. These manually crafted queries, derived from real-world reverse engineering efforts, reflect the practical requirements encountered during actual reverse engineering tasks. This approach underscores our commitment to ensuring that the evaluation closely mimics real-world scenarios. Finally, we obtain an evaluation set containing 257 queries and relevant function pairs.

\subsubsection{Baselines}

Previously in binary analysis research, the concept of assembly code search has not been explicitly proposed. Research efforts predominantly revolve around binary code similarity detection (BCSD), which only targets to compare the similarity between two assembly codes, instead of the natural language level functionality.
% , which aims to identify assembly code segments most likely compiled from an identical source code snippet, accounting for potential differences due to the compiler used and compilation options selected. 
% Therefore, as a baseline for our work, we use models trained directly from docstrings and compare the performance with several general embedding models. This includes sentence-transformers~\cite{} (which we use as the initial model for the text side), gte-large~\cite{liGeneralTextEmbeddings2023} (one of the best-performing open-source encoders on the METB leaderboard~\cite{}), SFR-Embedding-Mistral~\cite{wangImprovingTextEmbeddings2023,sfr-embedding-mistral} (the top-performing open-source model on the METB leaderboard, based on the mistral decoder architecture), and the vogaye-02 series models.
% Consequently, we use the models trained directly from extracted docstrings as our baseline.

\iffalse
Consequently, we use some general embedding models for comparison, including 
sentence-transformers
~\cite{reimersSentenceBERTSentenceEmbeddings2019,all-mpnet-base-v2}, the initialization model for the text encoder in Section~\ref{subsec:assembly-code-encoder-traininig}; 
GTE~\cite{liGeneralTextEmbeddings2023,gte-large}, recognized as one of the premier open-access encoders on the MTEB leaderboard
~\cite{muennighoffMTEBMassiveText2023,mteb-leaderboard}; 
% SFR-Embedding-Mistral
% ~\cite{wangImprovingTextEmbeddings2023}
% \footnote{\url{https://huggingface.co/Salesforce/SFR-Embedding-Mistral}}, the leading open-access model on the METB leaderboard which a finetuned Mistral decoder model~\cite{}; 
vogayeai models~\cite{voyage} and the OpenAI embedding models~\cite{openai,neelakantanTextCodeEmbeddings}.

\else

Consequently, we use some general embedding models for comparison, including 
sentence-transformers~\cite{reimersSentenceBERTSentenceEmbeddings2019}\footnote{\url{https://huggingface.co/sentence-transformers/all-mpnet-base-v2}}, the initialization model for the text encoder in Section~\ref{subsec:assembly-code-encoder-traininig};
GTE~\cite{liGeneralTextEmbeddings2023}\footnote{\url{{https://huggingface.co/thenlper/gte-large}}}, recognized as one of the premier open-access encoders on the MTEB leaderboard;
Voyage AI models\footnote{\url{{https://www.voyageai.com}}}, including the \texttt{voyage-code-2} model optimized for code retrieval;
and the OpenAI embedding models~\cite{neelakantanTextCodeEmbeddings}\footnote{\url{{https://openai.com/}}}.

\fi

\subsubsection{Metric}

% Therefore, we adopt an evaluation metric more commonly used in recommendation systems, namely Mean Average Precision (MAP).
During our manual reverse engineering process, we often encounter scenarios where a single function can be described in multiple ways, or multiple functions implement similar functionalities.
% In such situations, using recall as an evaluation metric may not be suitable due to the multiplicity of valid "correct" answers.
In such situations, the recall is defined as follows. Suppose $\{a_i\}$ is the set of relevant functions, and $\{b_i\}$ is the retrieved functions. And there are $q$ queries trying to retrieve the desired functions. The the $\text{recall}@k$ is defined as follow:

\begin{equation}
    \text{recall}@k = \frac{1}{q} \sum_{i = 1}^{q} \frac{|\{a_i\} \cap \{b_i\}|}{ \min({|\{a_i\}|, |\{b_i\}|})}
\end{equation}

when $|\{b_i\}| = k$.

% When the $|\{a_i\}| = 1$, that is there is only 
Moreover, we adopt MAP (Mean Average Precision) to better measure the ranking capability of the model, which is defined as:

\begin{equation}
    \text{AP} = \frac{1}{q} \sum_{k} P@k \times \text{rel}@k
\end{equation}

where $P@k$ refers to $\text{precision}@k$:

\begin{equation}
    P@k = \frac{|\{a_i\} \cap \{b_i\}|}{|\{b_i\}|}, \text{when} |\{b_i\}| = k
\end{equation}

and $\text{rel}@k$ is a relevance function, which is an indicator function that equals 1 if the function at rank k is relevant and equals 0 otherwise.

In the evaluations, the pool size is set to 10000, which means that the model needs to identify the relevant functions from 10000 functions.

\subsubsection{Main Result}

{\renewcommand{\arraystretch}{1.2}

\setlength\tabcolsep{3.5pt}

\begin{table}[t!]
\small
\centering
\begin{tabular}{l|ccccc}

% \Xhline{1pt}Model                 & Recall@1 & Recall@5 & Recall@10 & Recall@20 & MAP   \\ \hline
% sentence-transformers & 0.113    & 0.161    & 0.222     & 0.265     & 0.145 \\
% gte-large                   & 0.121    & 0.171    & 0.202     & 0.244     & 0.149 \\
% voyage-2$^*$              & 0.142    & 0.267    & 0.316     & 0.344     & 0.202 \\
% voyage-large-2$^*$        & 0.319    & 0.482    & 0.578     & 0.635     & 0.388 \\
% voyage-code-2$^*$         & 0.331    & 0.485    & 0.586     & 0.631     & 0.396 \\ 
% text-embedding-3-small$^\dagger$ & 0.173    & 0.278    & 0.351     & 0.406     & 0.225 \\
% text-embedding-3-large$^\dagger$ & 0.344    & 0.441    & 0.494     & 0.585     & 0.395 \\ \hline
% \docstring             & 0.323    & 0.522    & 0.593     & 0.644     & 0.412 \\
% \dsname         & \textbf{0.424}    & \textbf{0.567}    & \textbf{0.643}     & \textbf{0.723}     & \textbf{0.498} \\

\Xhline{1pt}Model                & Recall@1       & Recall@20      & MAP            \\ \hline
sentence-transformers            & 0.113          & 0.265          & 0.145          \\
gte-large                        & 0.121          & 0.244          & 0.149          \\
voyage-2$^*$                     & 0.142          & 0.344          & 0.202          \\
voyage-large-2$^*$               & 0.319          & 0.635          & 0.388          \\
voyage-code-2$^*$                & 0.331          & 0.631          & 0.396          \\
text-embedding-3-small$^\dagger$ & 0.173          & 0.406          & 0.225          \\
text-embedding-3-large$^\dagger$ & 0.344          & 0.585          & 0.395          \\ \hline
% \docstring                       & 0.323          & 0.644          & 0.412          \\
\dsname                          & 0.424 & 0.723 & 0.498 \\

\Xhline{1pt}
\end{tabular}
\caption{
\label{tab:overall-result}
The evaluation result of the model trained on \dsname, with the general encoders as baselines. $^*$Voyage AI embedding models. $^\dagger$OpenAI embedding models.
}
\end{table}
}

% In Table~\ref{tab:overall-result}, we display the comparison between our approach and the baseline models. It is evident that general models such as sentence-transformers, GTE, and voyage-2, which have not been exposed to substantial amounts of code during training – and even less so to assembly code – often struggle to grasp the control flow within assembly code. 

In Table~\ref{tab:overall-result}, our comparative analysis highlights the performance gap between our method and baseline models, notably sentence-transformers, GTE, and voyage-2. These general models tend to underperform in code search tasks involving assembly code due to their limited coding data during training, particularly the complex assembly code which needs an understanding of control flows. Their reliance on textual snippets solely within assembly code for semantic cues, missing the control flow information, often leads to suboptimal results in both recall and MAP metrics. In contrast, the voyage-code-2 and voyage-large-2 models demonstrate better performance, attributed to their more robust training on diverse code datasets compared to voyage-2.

% Interestingly, the performance of the model trained on the \docstring dataset—initiated from a basic model pre-trained with MLM task—matches up to that of the voyage-code-2 model, which benefits from training on a larger code dataset. This contrast suggests a substantial gap in the current code search models' ability to facilitate assembly code search effectively, as these models have not been sufficiently trained with extensive assembly code datasets to reach a functional level to support binary code analysis tasks. 

% Yet, the model's performance sees a significant enhancement upon incorporating code search datasets generated through the virtual compiler from various languages. This indicates that even though the virtual assembly code generated by the virtual compiler does not originate directly from a real compiler, it does not impede the model's semantic understanding. Therefore, enabling the encoder to encounter a wider array of semantic concepts through these virtually compiled datasets leads to improved code search capabilities.

Models trained on the \dsname dataset, starting from an MLM (Masked Language Modeling) pre-trained base model, still show significant improvements over OpenAI's models. This underscores a critical gap in the capability of existing code search models to effectively support assembly code search, indicating that these models lack comprehensive training with extensive assembly code datasets necessary to achieve a functional level for aiding binary code analysis.

\subsubsection{Virtual v.s. Real Assembly Code}

We analyze the effectiveness of virtual assembly code against actual assembly code in assembly code search tasks. By virtually compiling source code from \docstring into assembly code, we train two models using the real and virtual assembly code respectively, combined with docstrings, and compared their performance. The result is detailed in Table~\ref{tab:eval-virtualdocstring}, along with the result of excluding small functions from both sets of data (denoted as \texttt{w/o Small}), which follows our approach during the training phase in Section~\ref{subsec:data-processing}.

The performance analysis reveals that excluding small functions, both sets of assembly code demonstrate similar outcomes, indicating that the quality of virtual assembly code is on par with real assembly code. Not training the model on small functions is intended to reduce model hallucination, but this leads to minimal performance gains and poses a risk of generating irrelevant code when processing shorter source code. Conversely, the actual compiler-generated assembly for small functions aligns well with the corresponding docstrings, providing useful insights albeit not ideal for training purposes. This slight performance dip in the \docstring dataset without small functions underscores the nuanced impact of training data selection on model behavior.

{\renewcommand{\arraystretch}{1.2}
\begin{table}[t!]
\small
\centering
\begin{tabular}{l|ccc}
\Xhline{1pt}
Dataset              & Recall@1 & Recall@20 & MAP   \\
\hline
\docstring           & 0.323    & 0.644     & 0.412 \\
\quad w/o Small         & 0.329    & 0.612     & 0.402 \\
\virtualdocstring    & 0.315    & 0.620     & 0.393 \\
\quad w/o Small & 0.325    & 0.627     & 0.396 \\
\Xhline{1pt}
\end{tabular}
\caption{
\label{tab:eval-virtualdocstring}
Comparison of virtual assembly and the ground truth assembly code in assembly code search task.
}
\end{table}
}

\subsubsection{Dataset Contributions}

% In this section, we delve into the analysis of the contribution of each dataset utilized for assembly code search. Table~\ref{tab:eval-ingredient} shows the impact of different dataset components on the overall performance. 

{\renewcommand{\arraystretch}{1.2}
\begin{table}[t!]
\small
\centering
\begin{tabular}{l|ccccc}
\Xhline{1pt}Dataset       & Recall@1 & Recall@20 & MAP   \\ \hline
\vicds                    & 0.292    & 0.641     & 0.375 \\
\quad w/o \ \  Go         & 0.288    & 0.612     & 0.366 \\
\quad w/o \ \  Java       & 0.292    & 0.621     & 0.359 \\
\quad w/o \ \  Javascript & 0.296    & 0.618     & 0.364 \\
\quad w/o \ \  PHP        & 0.329    & 0.631     & 0.401 \\
\quad w/o \ \  Python     & 0.274    & 0.584     & 0.347 \\
\quad w/o \ \  Ruby       & 0.307    & 0.609     & 0.381 \\
\quad w/o \ \  C (CCSD)   & 0.272    & 0.578     & 0.348 \\ \hline
\docstring                & 0.323    & 0.644     & 0.412 \\ \hline
\dsname                   & 0.424    & 0.723     & 0.498 \\
\quad w/o \ \  PHP        & 0.418    & 0.701     & 0.495 \\
\Xhline{1pt}
\end{tabular}
\caption{
\label{tab:eval-ingredient}
The evaluation result of the contribution of different components. C is from the CCSD dataset.
}
\end{table}
}

This section evaluates how each dataset influences the assembly code search performance, as shown in Table~\ref{tab:eval-ingredient}. Training solely with the \vicds dataset underperformed compared to using the \docstring dataset. Detailed analysis of specific failures highlighted the representation challenge in assembly language. For instance, low-level operations like the "keccakf function" in SHA3 or "destroying a linked list" are broadly abstracted in higher-level languages through library calls, and are missing in broader datasets like CodeSearchNet. The absence of these concepts affects performance finally.

Further evaluations involve removing language components from the \vicds dataset, as detailed in Table~\ref{tab:eval-ingredient} with rows prefixed with \texttt{w/o}. Generally, omitting any single language dataset negatively impacts the results. Surprisingly, excluding the PHP dataset improved performance, likely due to suboptimal PHP handling by the virtual compiler, with the balanced performance in the \dsname dataset suggesting compensatory benefits from better-matched docstrings.

Additionally, the analysis identifies a significant drop in performance using the CodeSearchNet dataset alone, especially without \texttt{C (CCSD)}. This degradation highlights the importance of incorporating low-level concepts like "linked list" and "keccakf function" in training sets for assembly code search, even if these assembly codes are not compiled from a real compiler.

\section{Limitation \& Discussion}

\subsection{Code Search Dataset Quality}

The quality of the assembly code search dataset we construct is intrinsically tied to the underlying code search datasets, which are predominantly derived from docstrings, such as the CodeSearchNet dataset utilized in this study. In the evaluation in CodeSearchNet, 32.8\% of docstrings were found to be irrelevant to the source code. In our evaluation, a random sample of docstrings extracted from the Ubuntu source code exhibited a comparable proportion of loosely related docstrings. This observation underlines a significant limitation: relying solely on docstrings as the basis for code search datasets may introduce challenges due to the potential lack of strong relevance or association.
It's conceivable to generate code search datasets directly from source code using LLMs. 
% However, it's worth noting that the virtual compiler is still meaningful, with its capability to compile without bias and compile different programming languages.
However, with the capability to compile different programming languages, we can greatly augment the assembly code search dataset without bias.
% Because extracting source code and obtaining its corresponding assembly code is still a challenge, quite distinct from the straightforward extraction of source code alone. With a virtual compiler, we can greatly augment the assembly code search dataset without bias.

\subsection{Broader Impacts}

With the emergence of Large Language Models, code search has gained prominence as a crucial component within Retrieval-Augmented Generation (RAG)~\cite{langchain,gaoRetrievalAugmentedGenerationLarge2023}. 
With assembly code search as an RAG module, it helps the LLM to rapidly locate the desired functions with a large binary for faster reverse engineering.

While reverse engineering can be vital for security experts in vulnerability detection and analysis, it also poses risks of exploitation by malicious actors for software cracking and exploiting vulnerabilities. This duality highlights the necessity for ethical considerations and responsible use. 
% Advancing these tools should focus on enhancing security analysis capabilities while implementing safeguards and ethical guidelines to prevent misuse, ultimately aiming to bolster cybersecurity rather than undermine it.

% \subsection{Potential Risk of Reverse Engineering}

% In this paper, we have introduced numerous techniques related to reverse engineering. While reverse engineering can serve as a vital tool for security experts in vulnerability detection and security analysis, it also harbors the risk of being exploited by malicious actors for software cracking and vulnerability exploitation.

% This duality underscores the importance of ethical considerations and responsible use in the deployment of reverse engineering technologies. It is crucial that while advancing these tools to enhance security analysis capabilities, we also consider potential safeguards and ethical guidelines to prevent misuse. Developing these technologies with a focus on enhancing cybersecurity, rather than undermining it.

\subsection{Future Work}

We acknowledge that the evaluation dataset is relatively small, limited by the intensive labor and extensive analysis required for reverse engineers. Future work will focus on expanding and diversifying the evaluation dataset to enable a more comprehensive assessment of assembly code search.

In the model training in Section~\ref{subsec:assembly-code-encoder-traininig}, we train the assembly code model from scratch. However, with better model initialization from previous works such as CLAP~\cite{wang2024clap} and PalmTree~\cite{liPalmTreeLearningAssembly2021}, we can achieve better assembly code search performance and faster convergence.

\section{Conclusion}
In this study, we introduce a pioneering method to improve assembly code search by training an LLM to function as a virtual compiler, \sysname, effectively addressing the challenge of compiling difficulties and enhancing dataset quality for assembly code. This approach not only broadens the scope of languages compilable to assembly but also significantly boosts the performance of assembly code search tasks, outperforming all the existing solutions. Our results demonstrate the potential of virtual compilers to revolutionize reverse engineering processes, offering a promising direction for future advancements in software engineering and security analysis.

\section*{Acknowledgements}
We would like to sincerely thank all the reviewers for their valuable feedback that greatly helped us to improve this paper.
Additionally, special thanks are extended to Bolun Zhang and Jingwei Yi for their invaluable comments and assistance.

%% file: appendix.tex
\section{Illustration for Virtual Compiling non-C/C++ language}
\label{sec:illustration-virtual-compiling-golang}

Although our model training exclusively involved C/C++ code, the employment of Codellama—an extensively trained model across various programming languages—yields an unexpected and quite beneficial outcome. Rather than experiencing catastrophic forgetting, which is a common concern with LLMs as they undergo extensive continued training, our model retains its previously learned information. This retention facilitates the model’s understanding of other programming languages and enables it to generate corresponding assembly code effectively, akin to its behavior with C/C++.

Operating in this broader context without standard ground truth for these additional languages presented a unique challenge. To gauge the model's cross-language compilation capabilities, we sample some generated assembly code for these languages and manually check these samples to verify the correctness of the assembly output

% We show case an example in Section~\ref{sec:illustration-virtual-compiling-golang} in Appendix.

\begin{figure}[t]
    \centering
    \setlength{\abovecaptionskip}{2mm}
    \includegraphics[width=0.85\linewidth]{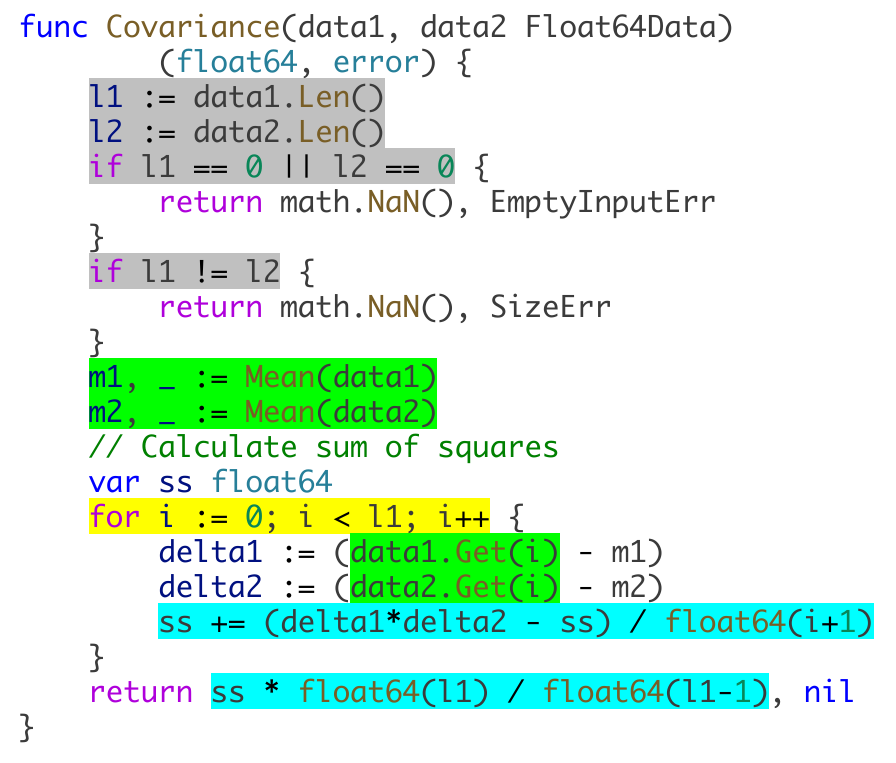}
    \caption{Example Golang function \texttt{Covariance} calculating the statistical covariance between two data sets, demonstrating input validation, mean calculation, and the sum of squares.}
    \label{fig:golang-source-code}
    % \vspace{-2mm}
\end{figure}

\begin{figure}[t]
    \centering
    \setlength{\abovecaptionskip}{2mm}
    \includegraphics[width=1\linewidth]{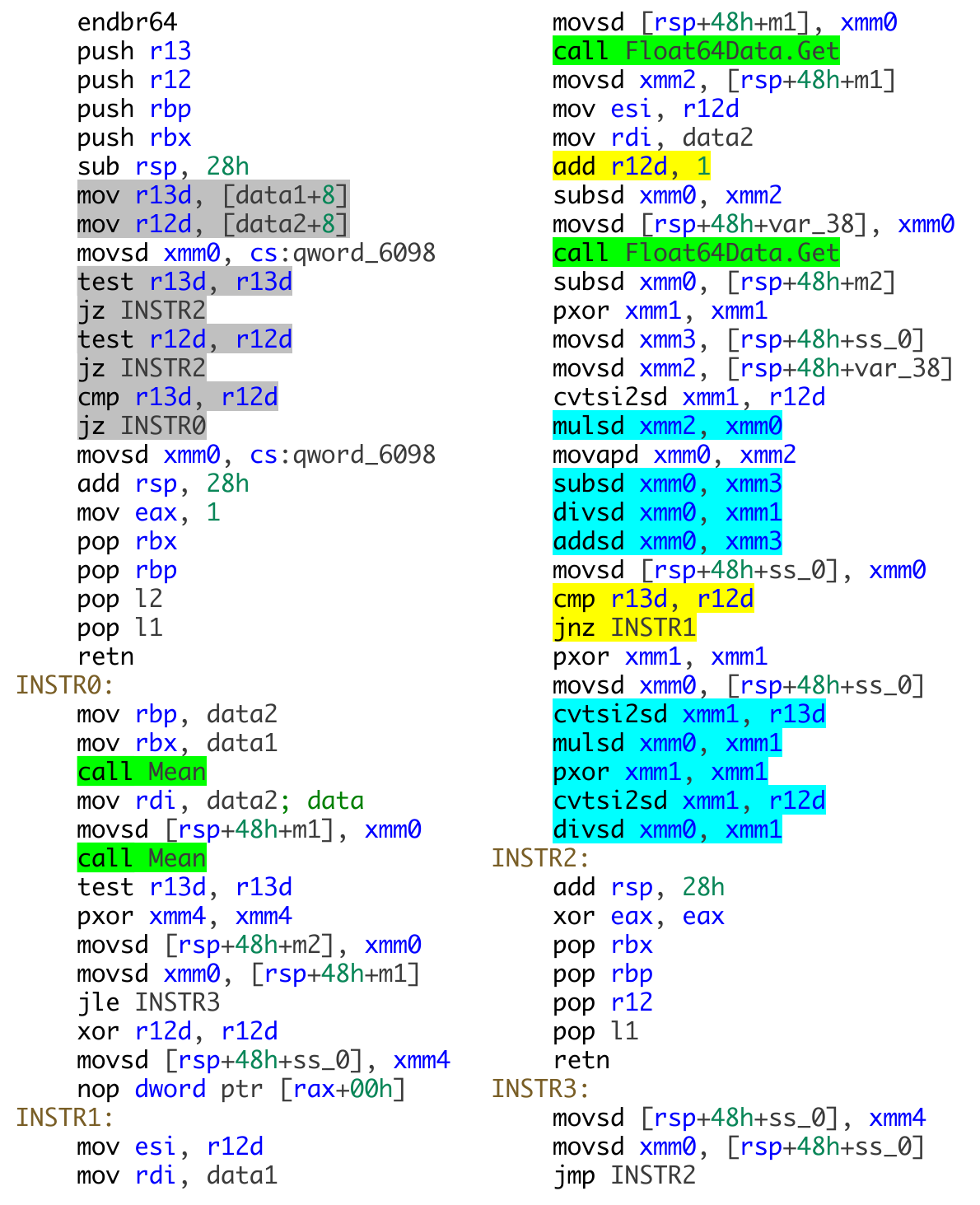}
    \caption{Virtual assembly code generated from the Golang \texttt{Covariance} function shown in Figure~\ref{fig:golang-source-code}, showcasing the translated computational logic and structure in low-level instructions. The corresponding function parts are highlighted in different colors.}
    \label{fig:golang-assembly}
    % \vspace{-2mm}
\end{figure}

% In this section, we qualitatively analyze the virtual compiler's proficiency in comprehending languages beyond C/C++ by showing a case. 
Figure~\ref{fig:golang-source-code} presents a Golang function for calculating covariance, while Figure~\ref{fig:golang-assembly} displays the corresponding virtual assembly code. The mapping between key elements of the source and assembly code is highlighted for clarity: the input validation segment of the function is marked with a grey background; function calls, including the Mean function and access to an item inside the Golang array, are illuminated in green; arithmetic operations performed during the covariance calculation are emphasized in cyan; and the for loop—encompassing boundary checks and the incrementation of the loop variable—is spotlighted in yellow.

\section{Docstring Cleaning}
\label{sec:docstring-clean}

Following a similar approach to the one used by CodeSearchNet for cleaning docstrings from function source code, we apply a series of cleaning steps to the docstrings we collected:

\begin{itemize}
    \item \textit{Eliminate docstring text borders}. We strip away leading and trailing `*' characters from docstrings, which act as decorative ``borders" around docstrings, and do not contribute to the semantic value of the query. 
    % This action is intended to focus queries on the core function description.
    \item \textit{Retain first paragraph}. Only the first paragraph of each docstring is kept to eliminate extensive details about function parameters and return values.
    \item \textit{Remove short docstrings}. After trimming down to the first paragraph, docstrings that are too short are considered as lacking meaningful content and, thus are removed.
\end{itemize}